\documentclass[useAMS,usenatbib, letters]{mn2e}

\usepackage{amsmath}
\usepackage{threeparttable}
\usepackage{graphicx, color}
\usepackage{bm}
\usepackage{soul}
\usepackage{amssymb}
\usepackage{hyperref}
\usepackage[utf8]{inputenc}

\newcommand{\eqb}{\begin{eqnarray}}
\newcommand{\eqe}{\end{eqnarray}}

\title[Pulsar Transits]{Radio Emission from Sgr~A*: Pulsar Transits Through the Accretion Disc}
\author[Christie et al.]
{I.~M. Christie$^{1}$\thanks{E-mail:ichristi@purdue.edu}, M.~Petropoulou$^1$\thanks{E-mail:mpetropo@purdue.edu; Einstein Fellow}, P.~Mimica$^{2}$\thanks{Email: petar.mimica@uv.es}, D.~Giannios$^{1}$\\ 
$^{1}$Department of Physics and Astronomy, Purdue University, 525 Northwestern Avenue, West Lafayette, IN 47907, USA \\
$^{2}$Departament d'Astronomia i Astrof\'{\i}sica Universitat de Val\`{e}ncia, Carrer del Dr. Moliner, 50, 46100 Burjassot, Valencia, Spain}

\begin{document}
\date{Received / Accepted}
\pagerange{\pageref{firstpage}--\pageref{lastpage}} \pubyear{2016}
\topmargin -0.5in
\maketitle
	
	\begin{abstract}
	Radiatively inefficient accretion flow models have been shown to accurately account for the spectrum and luminosity observed from Sgr~A* in the X-ray regime down to mm wavelengths. However, observations at a few GHz cannot be explained by thermal electrons alone but require the presence of an additional non-thermal particle population. Here, we propose a model for the origin of such a population in the accretion flow via means of a pulsar orbiting the supermassive black hole in our Galaxy. Interactions between the relativistic pulsar wind with the disc lead to the formation of a bow shock in the wind. During the pulsar's transit through the accretion disc, relativistic pairs, accelerated at the shock front, are injected into the disc. The radio-emitting particles are long-lived and remain within the disc long after the pulsar's transit. Periodic pulsar transits through the disc result in regular injection episodes of non-thermal particles. We show that for a pulsar with spin-down luminosity $L_{\rm sd} \sim 
3\times10^{35}$~erg~s$^{-1}$  and a wind Lorentz factor of $\gamma_{\rm w} \sim 10^4$ a quasi-steady synchrotron emission is established with luminosities in the $1-10$~GHz range comparable to the observed one. 	
	\end{abstract}
	\section{Introduction}
	\label{sec:introduction}
	
	The compact radio source Sgr~A*, associated with the supermassive black hole (SMBH) of mass $M_{\rm BH} = 4.3 \times 10^{6}$ M$_{\odot}$, marks the location of our Galactic Centre (GC) at a distance of $\sim 8.3$~kpc \citep{genzel2010, chatzopoulos2015}. Observations in the mm wavelength range up to the submillimeter bump can be described by the emission of thermal electrons in radiatively inefficient accretion flow (RIAF) models \citep{narayan1995_2}. However, a thermal distribution alone falls short of explaining radio observations in the few GHz range and requires an additional dominant contribution from non-thermal electrons in order to account for this deficiency \citep{ozel200,yuan2003}. An enhancement of the effective temperature within the disc by non-local transport processes can also be used in RIAF models in order to match the radio observations \citep{narayan1995_2}. Alternatively, the radio emission may originate in a outflow or jet \citep{markoff2007,falcke2009}. However, the evidence for the 
presence of a jet in Sgr~A* is inconclusive.
	
	The GC region is expected to harbor thousands of radio pulsars, but despite numerous searches, these objects remain elusive. Resolving the ``missing pulsar'' problem is both an observational and theoretical challenge. From an observational standpoint, these objects provide an excellent probe of the ionized gas surrounding Sgr~A* \citep{cordes1997} and can be used to test predictions of general relativity in the vicinity of the black hole \citep{pfahl2004}. A population of hundreds to thousands of observable radio pulsars have been proposed to reside within the inner pc of the GC \citep{wharton2012,chennamangalam2014}. A similar population size of millisecond pulsars (MSPs) was also argued to reside in the vicinity of the GC and can account for the \textit{Fermi} observed GeV excess \citep{brandt2015}. The view that a large number of young neutron stars are contained in the GC region has since been supported by the discovery of a magnetar located at a distance of $\sim0.1$ pc from the SMBH \citep{eatough2013,
mori2013,rea2013}.
	
	In this letter, we propose a mechanism for the injection of the non-thermal particles required to accurately describe the radio observations at a few GHz. A fiducial pulsar residing within $\sim0.1$ pc of the GC can interact with the accretion flow leading to the formation of a bow shock in the relativistic wind of the neutron star (for a derivation of its shape, see \cite{wilkin1996}, \cite{christie2016} and for an application to pulsar wind nebulae, see \cite{bucciantini2005}). Electron-positron pairs, accelerated at the shock front, mix with the turbulent disc. They remain in the disc for an accretion timescale while cooling via synchrotron radiation. The pulsar's orbit around the black hole results in regular interceptions with the disc which establishes a continuous injection of non-thermal electrons.\footnote{When stating electrons or particles, we are referring to both electrons and positrons.}
	
	This letter is structured as follows. In Section~\ref{sec:transit_properties}, we present the properties of the pulsar's transit through the disc.  In Section~\ref{sec:distribution_spectra}, we determine the temporal evolution of a distribution of non-thermal particles being injected within the disc and the resulting synchrotron spectrum. In Section~\ref{sec:orbits}, we study the cumulative emission resulting from continuous transits through the disc and conclude with a discussion in Section~\ref{sec:discussion}.
	
	\section{Pulsar Transits Properties}
	\label{sec:transit_properties}
\begin{figure}
			\centering
			\includegraphics[height=0.25\textwidth]{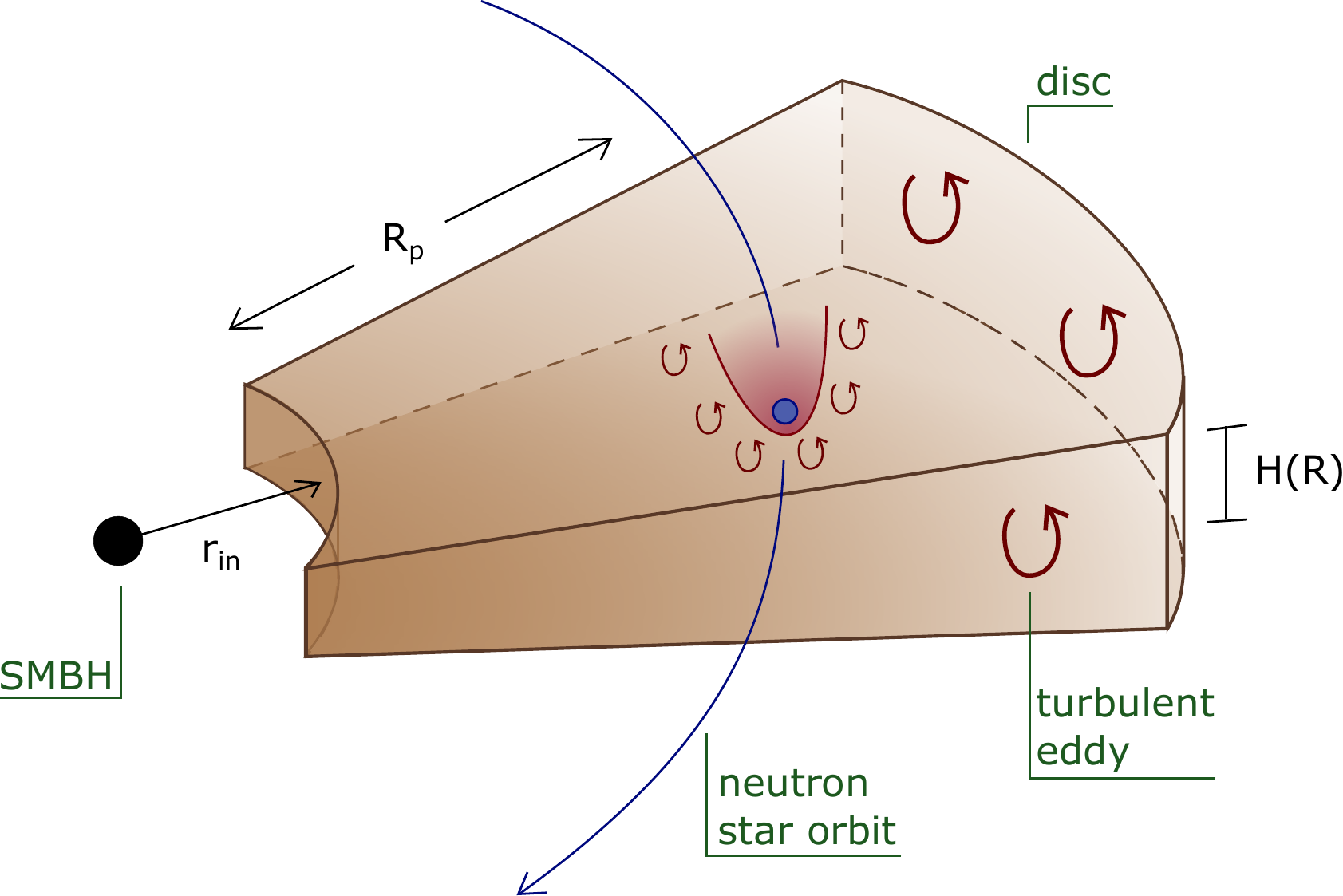}
			\caption{Sketch of a pulsar's transit through the accretion disc (not to scale). The pulsar is assumed to traverse a distance of $\sim 2 \, H(R_{\rm p})$ through the disc, where $H$ is the half-width of the disc. Relativistic electrons from the shocked pulsar wind, denoted by the shaded maroon region, spread throughout the disc mixing with the disc gas via turbulent eddies. The energetic particles remain in the disc until they accrete onto the black hole. While in the disc, these particles cool via synchrotron radiation.}
			\label{fig:pulsar_transit_sketch}
		\end{figure}

	From the discovery of the $0.1$ pc magnetar, \cite{giannios2016} argued that as many as $\sim 10$ neutron stars, with ages $\sim 10^{4}$~yr, may be orbiting Sgr~A*. Because of their young age, their spin-down luminosities can be very large $L_{\rm sd} \gtrsim 10^{35}$~erg~s$^{-1}$ \citep{manchester2005}. 
	We consider a pulsar whose orbit lies within 0.1~pc of the GC. For simplicity, we assume that the pericenter passage, at a radial distance $R_{\rm p}$ from the black hole, coincides with the pulsar's transit through the accretion disc of Sgr~A* (see Fig.~\ref{fig:pulsar_transit_sketch}). The characteristic stellar velocity at $R_{\rm p}$ is 
	\eqb
	\label{eqn:v_p}
	v_{\rm p} \simeq c \, \sqrt{\frac{2 \, R_{\rm g}}{R_{\rm p}}} \simeq 3.4\times10^8 \, R_{\rm p, 16}^{-1/2} \, {\rm cm} \, {\rm s}^{-1} 
	\eqe
	where  $R_{\rm g} = G \, M_{\rm BH}/c^{2} \simeq 6.4\times 10^{11}$~cm is the gravitational radius for a black hole of mass $M_{\rm BH} = 4.3 \times 10^{6}$~M$_{\odot}$. Henceforth, we adopt the notation $Q_{\rm x} = Q/10^{x}$ in cgs units. 
	The timescale in which it takes the pulsar to complete its pericenter transit through the accretion disc is estimated as  
	\eqb
	\label{eqn:t_p}
	t_{\rm p} \sim \frac{R_{\rm p}}{v_{\rm p}} \sim 3\times10^{7} \, R_{\rm p, 16}^{3/2} \, {\rm s} \sim 1 \, R_{\rm p, 16}^{3/2} \, {\rm yr}.
	\eqe 
	
	\textit{Chandra} can resolve the thermally emitting gas located at a radial distance of $R_{\rm b}\sim10^{17}$ cm where the inferred gas density is $n_{\rm b}=100$~cm$^{-3}$ \citep{baganoff2003}. For distances $<R_{\rm b}$, a geometrically thick accretion disc is expected to form \citep{roberts2016}. The density profile within the disc is model dependent, and can scale as $\propto R^{-3/2}$ for advection-dominated accretion flow (ADAF) models \citep{narayan1995} or $\propto R^{-1/2}$ for a convection-dominated accretion flow (CDAF) \citep{quataert2000}. Here, we adopt a density profile in the disc of $n(R)=n_{\rm b} \, (R_{\rm b}/R)$, such that $n_{\rm p} = 10^3 \, n_{\rm b, 2} \, R_{\rm b, 17} \, R_{\rm p, 16}^{-1}$ cm$^{-3}$. 
	The temperature of the disc is expected to approach the Virial temperature, regardless of the disc model. We may write the disc gas pressure as $P_{\rm th} =0.2 \, n_{\rm p} \, m_{\rm p} \, G \, M_{\rm BH} /  R_{\rm p}$, where $m_{\rm p}$ is the proton mass. The magnetic field pressure of the disc can be parameterized as a fraction $\epsilon_{\rm B}$ of the thermal pressure, thereby allowing us to estimate the magnetic field strength as $B_{\rm p} = \sqrt{8\pi \epsilon_{\rm B} P_{\rm th}} \simeq 0.007 \, \epsilon_{\rm B, -1}^{1/2} \, n_{\rm b, 2}^{1/2} \, R_{\rm b, 17}^{1/2} \, R_{\rm p, 16}^{-1}$~G.
	
	
	During the pulsar's transit through the disc, interactions between the pulsar wind and the disc lead to the formation of a termination shock in both mediums (i.e., a forward and a reverse shock in the disc and wind, respectively). The wind is terminated close to the pulsar, at a distance of $R_{\rm t} = \sqrt{L_{\rm sd}/4 \pi c m_{\rm p} n_{\rm p} v_{\rm p}^2} \sim 6 \times 10^{13} \, L_{\rm sd, 35.5}^{1/2} \, R_{\rm p, 16} n_{\rm b, 2}^{-1/2} R_{\rm b, 17}^{-1/2}$~cm \citep{giannios2016}. Relativistic particles, accelerated at the reverse shock, flow along with the shocked wind behind the star into a cylindrical tail region of typical size $\sim R_{\rm t}$, where the thermal pressure of the shocked fluid is $\sim P_{\rm th}$ \citep{bucciantini2001}. We assume that the accretion flow is turbulent i.e., it contains 
	eddies of different length scales $l$ that both disrupt the shocked wind and mix it with the disc gas. The mixing of the fluids due to eddies of length scale $l$ happens on their typical turnover timescale $t \sim t_{\rm p} (l/H(R_{\rm p}))^{2/3}$, where a Kolmogorov type of turbulent cascade was assumed  \citep{pope200} and $H$ is the half-width of the disc. The mixing on length scales of $R_{\rm t}$ happens fast compared to the pericenter time ($t \lesssim 0.1 \, t_{\rm p}$ for typical model parameters). The efficient mixing of the shocked wind with the disc gas allows the relativistic particles to be injected into the disc.
	
	The injected particles remain within the disc after the pericenter time until they accrete onto the black hole. While in the disc, these particles cool via synchrotron radiation. The accretion timescale is estimated as 
	$t_{\rm accr}(R_{\rm p}) \sim R_{\rm p}^{2}/ \alpha \, c_{\rm s}  H(R_{\rm p})$, where $H(R_{\rm p}) \sim R_{\rm p} \, (c_{\rm s}/v_{\rm p})$,   
	$c_{\rm s}\sim v_{\rm p} \, \sqrt{\Gamma/10}$ is the sound speed of the disc gas, $\Gamma$ is its adiabatic index,  and $\alpha$ is the disc $\alpha$-viscosity parameter. For $\Gamma = 5/3$ and $\alpha = 10^{-2} \, \alpha_{-2}$ the accretion timescale is 
	\eqb
	\label{eqn:t_accr}
	t_{\rm accr} \sim 600 \, t_{\rm p} \, \alpha_{-2}^{-1} \sim 600 \, R_{\rm p, 16}^{3/2} \, \alpha_{-2}^{-1} \,  {\rm yr}. 
	\eqe
	\begin{figure*}
		\centering
		\includegraphics[height=0.3\textwidth]{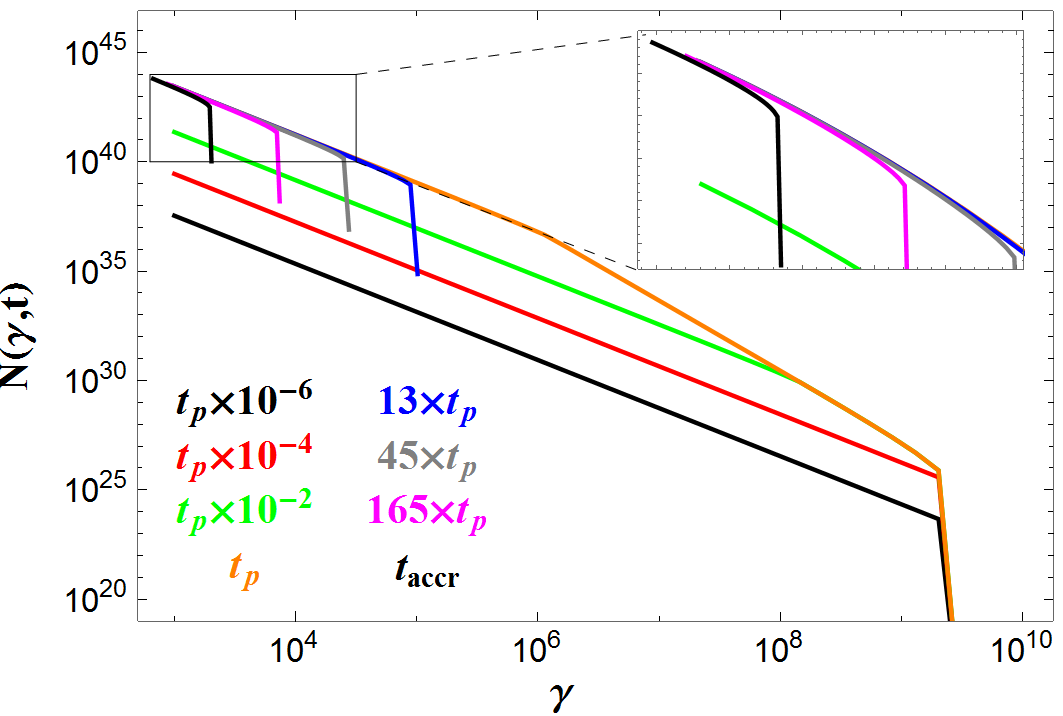}
		\includegraphics[height=0.3\textwidth]{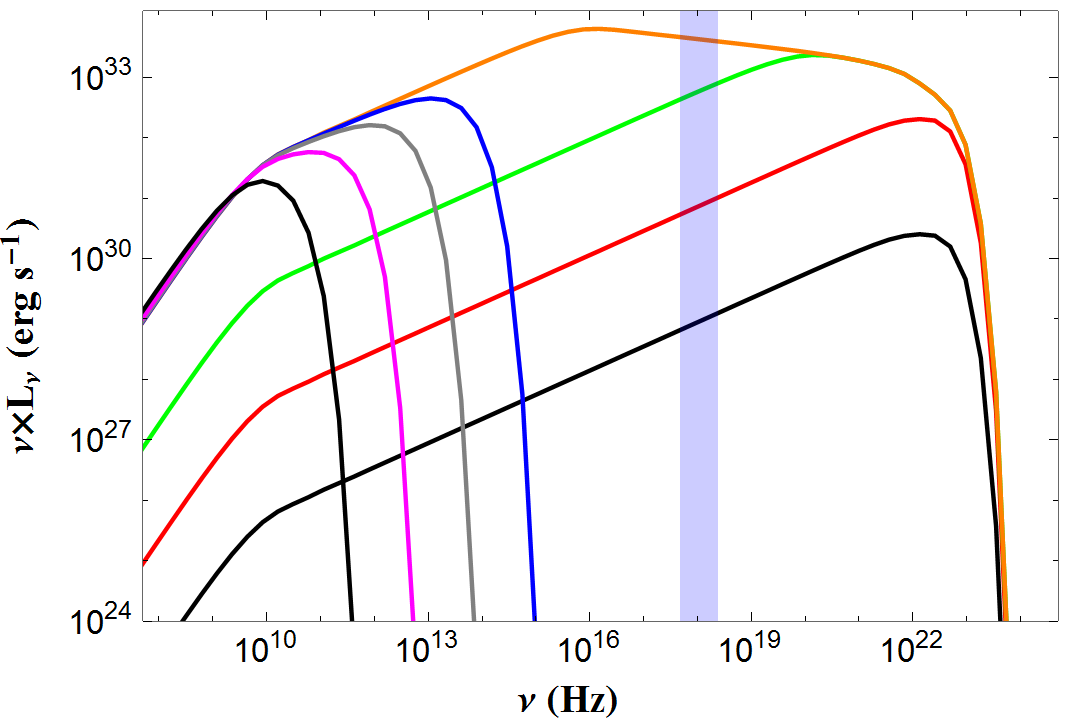}
		\caption{Plots of the temporal evolution of the electron distribution (left), as governed by eqns.~\ref{eqn:particle_distr_1} and \ref{eqn:particle_distr_2}, and the synchrotron spectra (right) for the times marked on the left plot. The shaded region on the right plot denotes the $2-10$~keV range.}
		\label{fig:particle_distribution_radiated_power}
	\end{figure*}
	
	\section{Particle Distribution \& Spectra}
	\label{sec:distribution_spectra}
	During the pericenter transit of the pulsar, relativistic electrons, assumed to follow a power-law distribution, are injected into the disc at a constant rate. The injection of particles ceases abruptly after the transit. These particles remain in the disc for an accretion timescale, during which they cool via synchrotron radiation. Let $N(\gamma, t)$ denote the number of electrons at time $t$ having Lorentz factors between $\gamma$ and $\gamma$+d$\gamma$. The temporal evolution of the particle distribution $N(\gamma, t)$ up to $t_{\rm p}$ is determined by the following equation
	\eqb
	\label{eqn:kinetic_equation}
	\partial_{\rm t} N(\gamma,t) - b \, \partial_{\gamma} [\gamma^{2} \, N(\gamma,t)] = Q_{\rm e}(\gamma, t),
	\eqe
	with the injection rate of relativistic particles being
	\eqb
	Q_{\rm e}(\gamma,t) = Q_0 \, \gamma^{-p} \, S(\gamma; \gamma_{\rm min}, \gamma_{\rm max}) \, S(t; 0, t_{\rm p}).
	\eqe
	Here, $S(y;y_1 ,y_2 )$ is a unit boxcar function, $b\equiv B_{\rm p}^{2} \sigma_{\rm T}/6 \pi m_{\rm e} c$, and $Q_0$ is 
	\eqb
	\label{eqn:Q_0}
	Q_0 = \frac{L_{\rm sd} (p-2)}{m_{\rm e} c^{2} (\gamma_{\rm min}^{2-p} - \gamma_{\rm max}^{2-p})} \quad ; \, p\ne2.
	\eqe
	Assuming Bohm acceleration at the reverse shock, the maximum Lorentz factor of the particles is $\gamma_{\rm max} \approx \sqrt{6 \pi q/B_{\rm p} \sigma_{\rm T}} \approx 1.4\times10^9 \, \epsilon_{\rm B,-1}^{-1/4} \, n_{\rm b, 2}^{-1/4} \, R_{\rm b,17}^{-1/4} \,R_{\rm p, 16}^{1/2}$. The adopted $\gamma_{\rm min}$ value corresponds to a Lorentz factor in the un-shocked pulsar wind of $\gamma_{\rm w} = \gamma_{\rm min} \, (p-1)/(p-2) = 6 \, \gamma_{\rm min}$, assuming $p=2.2$.
		\begin{figure*}
			\centering
			\includegraphics[height=0.3\textwidth]{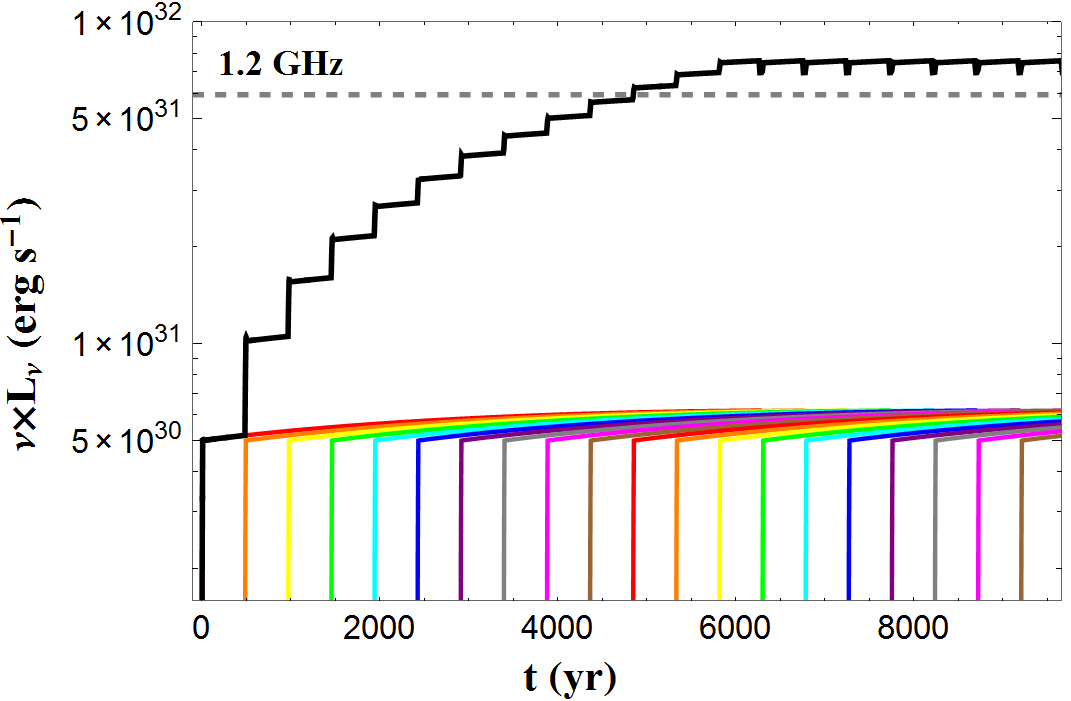}
			\includegraphics[height=0.3\textwidth]{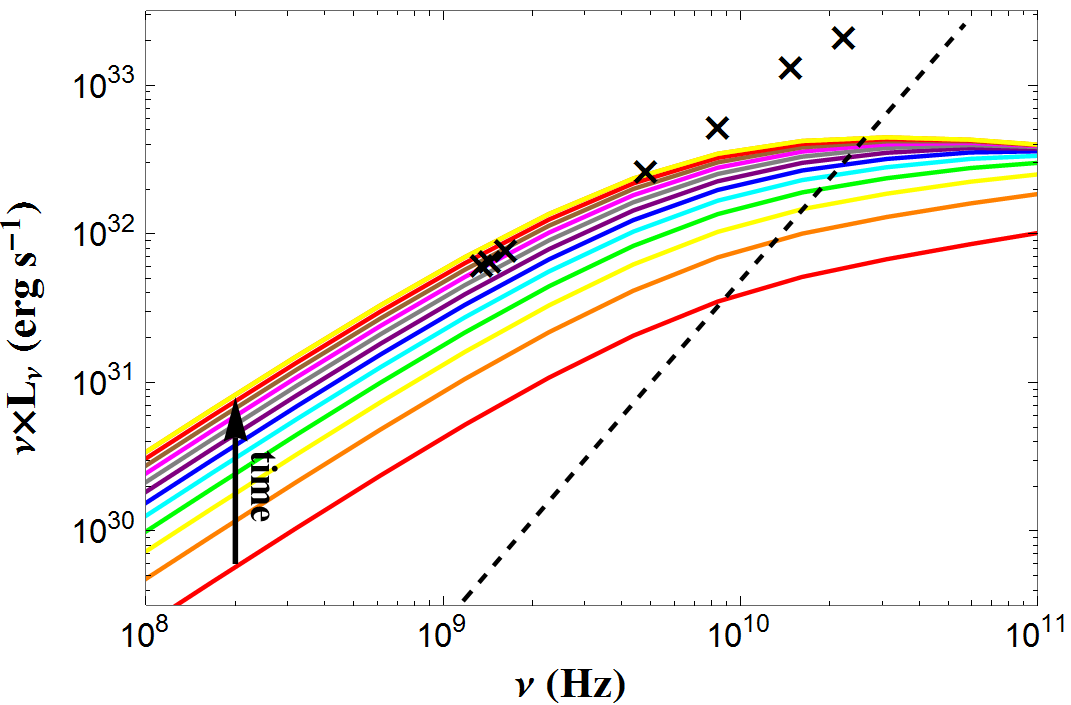}
			\caption{\textbf{Left:} Plot of the cumulative light curve at $1.2$ GHz for twenty pericenter transits through the accretion disc, shown as individual colored lines. The horizontal, dashed line denotes the typical 1.2~GHz luminosity from Sgr~A*. \textbf{Right:} Plot of the temporal evolution of the synchrotron spectrum overplotted with the observed radio spectrum \citep{falcke1998}. This model can account for the observed emission $\lesssim 10^{10}$~Hz. For higher frequencies, cyclo-synchrotron emission from thermal electrons, represented by the dashed line adopted from \protect\cite{ozel200}, is expected to dominate.}
			\label{fig:cumulative_emission_spectra}
		\end{figure*}
		
	Eqn.~\ref{eqn:kinetic_equation} can be solved analytically \citep{kardashev1962} and has the following solution 
	\eqb
	\label{eqn:particle_distr_1}
	N_{<t_{\rm p}}(\gamma, t) = \frac{Q_0 \, \gamma^{-1-p}}{b \, (p-1) } \, f(\gamma, t)
	\eqe
	where the subscript ``$< \, t_{\rm p}$'' refers to times below $t_{\rm p}$ and $f(\gamma,t)$ is defined as
	\eqb
	\label{eqn:piecewise}
	f(\gamma, t)=  \begin{cases}
		1 - (1 - b \, t \, \gamma)^{p-1} & \gamma_{\rm min} < \gamma \leq \gamma_{\rm c1}(t) \\
		1 - (\frac{\gamma_{\rm max}}{\gamma})^{1-p} & \gamma_{\rm c1}(t) \leq \gamma < \gamma_{\rm max}\\
		[1 - (\frac{\gamma_{\rm max}}{\gamma_{\rm min}})^{1-p}](\frac{\gamma_{\rm min}}{\gamma})^{1-p} & \gamma_{\rm c1}(t) < \gamma \leq \gamma_{\rm min}\\
		[(\frac{\gamma_{\rm min}}{\gamma})^{1-p} - (1 - b \, t \, \gamma)^{p-1}] & \gamma_{\rm c2}(t) \leq \gamma < \gamma_{\rm min}
	\end{cases},
	\eqe
	where $\gamma_{\rm c1}(t) \equiv \gamma_{\rm max}/(1 + b \, \gamma_{\rm max} t)$ and 	$\gamma_{\rm c2}(t) \equiv \gamma_{\rm min}/(1 + b \, \gamma_{\rm min} t)$.
	
	The injection of accelerated particles ceases for $t > t_{\rm p}$ at which point the pulsar has left the disc. However, the particles from the initial injection remain in the disc and continue to cool as they are advected radially inwards. To track the temporal evolution of the remaining particles for $t > t_{\rm p}$, we solve eqn.~\ref{eqn:kinetic_equation} without an injection term (i.e. $Q_{\rm e}(\gamma, t) = 0$) with the condition that $N_{>t_{\rm p}}(\gamma, t_{\rm p}) = N_{<t_{\rm p}}(\gamma, t_{\rm p})$. The resulting solution \citep{kardashev1962} is
	\eqb
	\label{eqn:particle_distr_2}
	N_{>t_{\rm p}}(\gamma, t) = (1 - b \, \gamma \, (t - t_{\rm p}))^{-2} \, N_{<t_{\rm p}}\left(\frac{\gamma}{1-b \, \gamma \, (t-t_{\rm p})}, t_{\rm p}\right).
	\eqe
	
	For the numerical examples presented in this letter, we adopt the following parameter set: $\gamma_{\rm min} = 10^3 \, \gamma_{\rm min, 3}$, $L_{\rm sd} = 3\times10^{35} \, L_{\rm sd, 35.5}$ erg s$^{-1}$, $p=2.2$, and $R_{\rm p, 16} = 5$. As the pulsar properties cannot yet be constrained, we provide general expressions for the parametric dependence of our results. The temporal evolution of the particle distribution is shown by the left panel in Fig.~\ref{fig:particle_distribution_radiated_power}. Shortly after $t_{\rm p}$, particles with high Lorentz factor quickly cool resulting in a narrower $\gamma$ range for the distribution (see inset plot in left panel of Fig.~\ref{fig:particle_distribution_radiated_power}). 
	
	The total radiated power per unit frequency for a distribution of relativistic electrons is calculated using the standard equations \citep{rybicki1986}.
	The temporal evolution of the synchrotron spectrum is presented in the right panel of Fig.~\ref{fig:particle_distribution_radiated_power}. There is a steady increase in luminosities over all frequencies up to $t_{\rm p}$ after which all emission above the near infrared (IR) regime quickly decreases. This sudden decrease is due to both fast cooling electrons and the cessation of freshly injected particles. 
	
	The X-ray luminosity at $t_{\rm p}$ (shaded region in the right panel of Fig.~\ref{fig:particle_distribution_radiated_power}) is given by
	\eqb
	\label{eqn:x_ray_lum}
	\nu L_{\nu} \sim 4 \times 10^{33} \, L_{\rm sd, 35.5} \, \gamma_{\rm min, 3}^{1/5} \, \frac{ \epsilon_{\rm B, -1}^{1/20}n_{\rm b, 2}^{1/20}\, R_{\rm b, 17}^{1/20}}{R_{\rm p, 16}^{1/10} \nu_{18}^{1/10}} \, {\rm erg} \, {\rm s}^{-1},
	\eqe
	where we have used $p=2.2$. As long as the X-ray emitting electrons are fast cooling, the X-ray luminosity is proportional to $L_{\rm sd}$ with a very weak dependence on all other parameters, including the density profile of the disc. The short synchrotron cooling timescale for the X-ray emitting electrons leads to a sharp drop in the emission after the transit. We therefore predict a non-thermal X-ray flaring event, with luminosities given by eqn.~\ref{eqn:x_ray_lum} and duration determined by the pericenter time (see eqn. (\ref{eqn:t_p}) and \citet{giannios2016}). The quiescent X-ray luminosity observed from the GC is measured at $L_{\rm X} \approx 2.4 \times 10^{33}$~erg~s$^{-1}$ \citep{baganoff2003}, making the emission from such a transit, in principle, detectable. This flaring event should not be confused with short duration IR and X-ray flares, occurring on timescales of minutes to hours, emanating from the GC \citep{baganoff2003,ghez2004}.

	The radio-emitting electrons have a long cooling timescale, $t_{\rm cool} \gg t_{\rm p}$, and are therefore left in the disc to radiate long after the transit. Our model predictions for the radio emission are not sensitive to the acceleration mechanism, in contrast with the X-ray emission, and fall within the $\nu^{1/3}$ part of the spectrum. 
	
	\section{Multiple Transits Through Disc}
	\label{sec:orbits}
	Members of the S-Cluster, a group of tens of massive stars observed within 0.1 pc of the GC, are characterized by highly elliptical orbits \citep{gillessen2009}. Here, we consider a pulsar with similar orbital parameters, i.e. with an apocenter distance of $R_{\rm apo}\approx5 R_{\rm p}\approx 2.5 \times 10^{17} \, R_{\rm p, 16.7}$~cm, which corresponds to an orbital period of $T_{\rm orb}\approx485$~yr. 
	The number of transits through disc on one accretion timescale is estimated as $N_{\rm orb} \sim t_{\rm accr}(R_{\rm p})/T_{\rm orb} \sim 13$, where we assumed one pulsar transit per orbital period.

	During each of these transits through the disc, the pulsar continuously injects more particles thereby increasing the cumulative emission. It takes $\sim N_{\rm orb}$ transits for the establishment of a quasi-steady emission, as shown by the $1.2$~GHz light curve in the left panel of Fig.~\ref{fig:cumulative_emission_spectra}. The resulting emission reaches a luminosity that is comparable to the observed one (represented by the horizontal, dashed line). 
	
	The temporal evolution of the radio synchrotron spectrum over a period of 13 transits, overplotted with GC radio observations \citep{falcke1998}, is shown in the right panel of Fig.~\ref{fig:cumulative_emission_spectra}. Each successive line represents the cumulative spectrum after an additional transit through the disc. In the GHz frequency range, the model-predicted synchrotron spectrum is comparable with the observed one in terms of spectral shape and luminosity.
	
	The dependence of the quasi-steady radio emission on the selected parameters can be estimated as follows
	\eqb
	\label{eqn:scaling_eqn_1}
	\nu L_{\nu} \approx N_{\rm tot} \, N_{\rm orb} \, (\nu P_{\nu}),
	\eqe
	where $N_{\rm tot} \sim L_{\rm sd} \, t_{\rm p} / m_{\rm e} c^2 \gamma_{\rm w}$ is an estimate for the total number of injected particles and $\nu P_{\nu}$ is the total radiated power per particle. The latter component has asymptotic analytical expressions for the $\nu^{1/3}$ part of the synchrotron spectrum (see \cite{rybicki1986}). The cumulative quasi-steady emission at 1 GHz is
	\eqb
	\label{eqn:scaling_eqn_2}
	\begin{split}
	\nu  L_{\nu} \sim 8 \times10^{31} \, L_{\rm sd, 35.5} \, \alpha_{-2}^{-1} \, \nu_{9}^{4/3} \, \gamma_{\rm w, 3}^{-1} \, \gamma_{\rm min, 3}^{-2/3} \, \times \\ \epsilon_{\rm B, -1}^{1/3} n_{\rm b, 2}^{1/3} R_{\rm b, 17}^{1/3}
	R_{\rm p, 16}^{5/6} \left(1 + \frac{R_{\rm apo, 17.4}}{R_{\rm p, 16}}\right)^{-3/2} {\rm erg} \, {\rm s}^{-1}.
	\end{split}
	\eqe
	This depends weakly on the magnetic field strength and, in turn, on the disc density at the pericenter distance. This approximation for the cumulative emission can be applied to obtain an estimate for the quasi-steady radio emission using other pulsar parameters with the requirement that the considered frequency is $\nu < \nu_{\rm c}$ and $\nu < \nu_{\rm min}$. Here, $\nu_{\rm c}$ and $\nu_{\rm min}$ are the synchrotron frequencies corresponding to Lorentz factors of electrons that cool over $t_{\rm accr}$ and that are injected at $\gamma_{\rm min}$, respectively. 
	
	\section{Discussion}
	\label{sec:discussion}	
	The injection of non-thermal particles, required in RIAF models to explain the radio emission at a few GHz, can result from a pulsar's transit through the accretion disc of Sgr~A*. We consider the pulsar's orbit to lie within $\sim0.1$ pc from the GC and track the continuous injection of particles over each transit. We find that a quasi-steady emission is established with luminosities in the GHz range comparable to observations. Our fiducial pulsar was chosen to have a spin-down luminosity of $L_{\rm sd} = 3\times 10^{35}$~erg~s$^{-1}$, which is typical for young neutrons stars with ages $\lesssim 10^{5}$~yr. Since a magnetar with an age less than this is known to reside in the vicinity of the GC and magnetars are a modest fraction of young neutron stars, choosing such a value for the luminosity is not extreme.	
	Similar results for the cumulative radio emission can come from the transits of multiple pulsars. However, the pulsar with the largest spin-down luminosity is likely to dominate the cumulative emission.
	
	In this Letter, we focused on the emission from particles accelerated at the reverse shock. A low Mach number shock is also expected to form in the disc. However, its contribution to the cumulative synchrotron emission is expected to be negligible, as the injected power in accelerated electrons is much lower than $L_{\rm sd}$. It is noteworthy that the GHz emission produced at the reverse shock does not depend on the details of the acceleration process, as it is produced by the pairs thermalized at the shock. These particles dominate in both number and energy downstream from the shock front \citep{sironi2009}. 
	
	Synchrotron self-absorption is not important in the observed bands. Considering a distribution of energetic particles residing within a region with volume $\sim R_{\rm p}^{3}$ the optical depth was determined to be $\tau \ll 1$ for frequencies $\nu \geq 10^{8}$~Hz, therefore allowing these effects to be ignored within our model.
	
	Throughout this work, the magnetic field strength of the disc was taken to be constant. It is expected, however, that the magnetic field is radially dependent and, more specifically, increases in the inner disc regions \citep{sadowski2013}. Considering this would lead to a more refined cumulative radio emission produced by the pairs. The presence of a small non-thermal population of electrons has also been shown to increase both the effective brightness temperature of the disk as well as the shape of the image of the source \citep{ozel200}. Ultimately, an advancement upon our model can be made by carefully following the pairs as they are advected through the disc.
	
	The detection of GC pulsars is required to evaluate the feasibility of our model and to constrain its parameters. \cite{rajwade2016} argue that previous surveys have only probed $\sim 2\%$ of the total pulsar population within 1~pc of GC. Future surveys with the Square Kilometre array (SKA), with initial operations beginning in 2020, would probe a large portion of the proposed population. 
	
	\section*{Acknowledgments}
	We thank the anonymous referee for insightful comments and Dr.~S.~Dimitrakoudis for producing the sketch presented in Fig.~\ref{fig:pulsar_transit_sketch}.
	M.P. acknowledges support for this work by NASA through Einstein Postdoctoral
	Fellowship grant number PF3~140113 awarded by the Chandra X-ray
	Center, which is operated by the Smithsonian Astrophysical Observatory
	for NASA under contract NAS8-03060. PM acknowledges the support from the European Research Council (grant CAMAP-259276), and the partial support of grants AYA2015-66899-C2-1-P and PROMETEO-II-2014-069.

	\bibliographystyle{mn2e} 
	\bibliography{sample}
\end{document}